 \documentclass[%
 reprint,
 twocloumn,
 amsmath,amssymb,
 aps,
]{revtex4-2}
\usepackage[a4paper, margin=1in]{geometry}
\usepackage[utf8]{inputenc}
\usepackage{textcmds}
\usepackage{xcolor}
\usepackage{dsfont}
\usepackage{amsmath}
\usepackage{amsthm}
\usepackage{amsfonts}
\usepackage{amssymb}
\usepackage{bm}
\usepackage{dsfont}
\usepackage{algorithm}
\usepackage{algpseudocode}
\algrenewcommand\algorithmicuntil{\textbf{within}}
\usepackage{exscale}
\usepackage{relsize}
\usepackage{upgreek}
\usepackage{lmodern}
\usepackage{graphicx}
\usepackage{floatflt}
\usepackage{here}
\usepackage{enumitem}
\usepackage{hyperref}
\usepackage{caption}
\usepackage{subcaption}
\renewcommand{\vec}[1]{\pmb{#1}}

\newcommand{\add}[1]{\textrm{{\textcolor{black}{#1}}}}

\begin{document}

\title{Properties of stable ensembles of Euclidean random matrices}
    \author{Philipp Baumgärtel}
    \author{Florian Vogel}
    \author{Matthias Fuchs}
	\affiliation{Fachbereich Physik, Universität Konstanz, 78457 Konstanz, Germany}

	\begin{abstract}
		 We study the spectrum of a system of coupled disordered harmonic oscillators in the thermodynamic limit. This  Euclidean random matrix ensemble has been suggested as model for the low-temperature vibrational properties of glass.   Exact numerical diagonalization is performed  in three and two spatial dimensions\add{, which is accompanied by a detailed finite size analysis}. It reveals a low-frequency regime of sound waves that are damped by Rayleigh scattering. At large frequencies localized modes exist. In between, the central peak in the vibrational density of states is well described by Wigner's semicircle law  \add{for not too large disorder, as is} expected for simple random matrix systems. We compare our results with predictions from two recent self-consistent field theories.
	\end{abstract}

	\date{\today} 
		\maketitle 

\section{Introduction}

The nature of the vibrational excitations in athermal amorphous solids remains an open and important topic affecting inter alia thermal properties of glasses at low temperatures \cite{Binder}. The vibrational properties of glassy materials differ strongly from the ones of crystals \cite{Schirmacher_Heterogeneous_Elasticity} as is well established by scattering  experiments \cite{Monaco2009,Baldi2010,Baldi2013,Baldi2014, PhysRevE.52.4026}. Computer simulations of various particle models have shown that the preparation of the glass state matters and a number of diverse phenomena have been discovered \cite{Mizuno2014,Gelin2016,Shimada2018,Mizuno2017,Wang2019,Wang2019-2,Saitoh2021,Hu2022,Szamel2022, PhysRevLett.127.215504, Horbach2001HighFS, Shimada2019VibrationalSD,Kapteijns2021}. 
As a consistent comprehensive theoretical picture is still lacking \cite{Lerner2021},  a simple idealized model appears desirable, where a part of the phenomena can be studied in detail. 

In 1999, Mezard, Parisi, and Zee introduced ensembles of Euclidean random matrices (ERM) \cite{Mezard1999}, which were studied as simple models for low temperature glasses by Grigera and colleagues \cite{Martin-Mayor2001,Grigera2001,Ciliberti2003,Grigera2011}, and by Schirmacher and colleagues \cite{Ganter2010,Ganter2011,Folli2017,Schirmacher2019}. Random matrix ensembles have successfully been employed in a wide variety of physical systems with disorder \cite{Mehta2014,Skipetrov2013, PhysRevB.103.104204}, and the special ERM ensemble may arguably be considered the most idealized model for the vibrations in glass. {\em Particles perform harmonic motion around random positions.  The restoring forces depend on the distances between the  positions via a positive spring function, and translational invariance is postulated. Only properties averaged over the random positions are studied.}
Based on diagrammatic perturbation expansions in field theoretic approaches, self-consistent theories for the central Greens functions have been developed \cite{Grigera2001,Ciliberti2003,Ganter2010,Ganter2011,Vogel2023}. They allow the (approximate) calculation of the vibrational density of states (vDOS) and of the dynamical structure factor. 

While ERM had been studied intensively up to a decade ago,  
open questions remained on the sound damping and on the spatial characteristics of the eigenmodes. In the present contribution, we investigate the most simple  ERM model, already considered in \cite{Grigera2001, Ganter2011}, an homogeneous and isotropic system with a positive Gaussian spring function\add{, and resolve the open questions: Even though the ERM system is purely harmonic, Rayleigh damping of sound arises because plane waves are not precise eigenmodes of the Hessian}. The only state parameter turns out to be the rescaled density $n$, which encodes the amount of disorder. We perform large scale numerical investigations including studies of finite size corrections, in order to reveal the complete characteristics of this specific ERM ensemble \add{for not too small $n$}.   We also compare with predictions from two self-consistent theories \cite{Grigera2001,Vogel2023}, where different series of diagrams in the perturbation expansion \add{in $1/n$} \cite{Grigera2011} were re-summed. 

\section{Model}
In this work we study an Euclidean random matrix (ERM) model \cite{Mezard1999,Skipetrov2013,Schirmacher2019} in which we consider $N$ particles which are randomly placed in a box of Volume $V = L^d$. Here, $d$ denotes the dimension of the system, and $L$ is its length. We apply periodic boundary conditions to the system and study an uniform distribution of particles. The set of random positions $\{\vec{r}_i\}$ will be called inherent positions. We consider a harmonic motion of the particles around their inherent positions which leads us to the definition of a random matrix $\vec{M}$ via the interaction potential $U$
\begin{align}
U(\vec{\phi}) &=    \frac{1}{4}\sum_{i,j}f(\vec{r}_i-\vec{r}_j)(\phi_i - \phi_j)^2 \notag\\ &= \frac 12 \sum_{i,j} M_{ij}\phi_i\phi_j,
\end{align}
with
\begin{align}
    M_{ij} = -f(\vec{r}_i-\vec{r}_j) +  \delta_{ij}\sum_kf(\vec{r}_i-\vec{r_k}).
\end{align}
Here, $f(\vec{r})$ is called spring function and $\phi_i$ is a small scalar displacement of particle $i$  from its inherent site.  \add{The scalar displacements mimick transverse displacements as coupling to density is neglected.}
In this work we consider the simple case, where the spring function is isotropic and given by the Gaussian 
\begin{align}\label{eq:springFunction}
    f(r) = \exp(-r^2/2),
\end{align}
with $r$ the dimensionless distance.
For positive spring functions, $f(r)>0$,  the potential $U$ is positive and thus  the matrix $\bf M$ is positive semi-definite.
In the limit of large systems, a single state parameter $n=N/V$ determines the properties of quantities averaged over the disorder.

In the harmonic approximation, the equations of motion of the system are given by
\begin{align}\label{EOM}
    \ddot{\phi}_i = -\sum_{j=1}^N  M_{ij}\phi_j \quad, \; \mbox{for } 1\le i \le N
\end{align}

Translational invariance and hence momentum conservation follow immediately from the potential $U(\vec{\phi})$.  Consequently, $\bf M$ has the eigenvalue zero. The associated eigenvector $\vec{e}^0$ corresponds to the uniform shift  $\vec{  e}^0=(1,1,....,1)/\sqrt{N}$.\\
Note that we already set the length scale of our system to one by the definition of the spring function in Eq.~\eqref{eq:springFunction},  and frequency and time are also chosen dimensionless quantities in Eq.~\eqref{EOM}.

\section{Methods}
We use two methods to study the characteristics of the system.  The first one, in which we diagonalize the random matrix $M_{ij}$, will be called normal mode analysis. The second one, where we solve the equations of motion, will be called excited wave analysis. \add{While the first method provides information  in frequency space (such as the density of states), the second provides the temporal evolution,  which is more revealing on the damping.} In both cases, averages over the disorder are finally performed by sampling different inherent positions. 

\subsection{Normal mode analysis}
In the normal mode analysis~\cite{Wang2019,Mizuno2014,Monaco2009,Shimada2018,Mizuno2017,Wang2019-2,Saitoh2021}, we calculate the eigenvalues $\lambda^k$, corresponding to the eigenfrequencies $\omega^k = \sqrt{\lambda^k}$, and the eigenvectors $\vec{e}^k$ of the random matrix $\vec{M}$. For this, we use the standard diagonalization routine of \textit{matlab} and a routine called lobpcg~\cite{lobpcg} which can handle sparse matrices efficiently.
Note, that the symmetry and semi-positivity of  $\vec{M}$ assures $\lambda^k\ge 0$ and that the $\vec{e}^k$ form an orthonormal basis.

The density of states per particle in the energy domain is calculated by \cite{Ashcroft76}
\begin{align}
    g_\lambda(\lambda) = \frac{1}{N}\overline{\sum_k\delta(\lambda-\lambda^k)}
\end{align}
and can be transformed into the frequency domain with $\lambda=\omega^2$ leading to
\begin{align}
    g(\omega) = 2\omega\; g_\lambda(\lambda(\omega)).
\end{align}
Here, the overbar denotes the average over disorder. As we expect discrete eigenfrequencies in finite systems, the density of states $g(\omega)$ can be subject to inaccuracies occurring due to the binning process if the bin size is chosen incorrectly. A quantity which resolves this issue is called integrated density of states 
 \cite{Wang2021,Wang2022,Wang2023} and can be calculated by
\begin{align}
    I(\omega) = \int_0^\omega g(\omega')\; \mathrm{d}\omega'.
\end{align}
The integrated density of states counts the number of eigenfrequencies up to a frequency $\omega$.\\

We calculate the dynamical structure factor by $S(q,\omega) = 2\omega\,S_\lambda(p,\lambda)$, where
\begin{align}
    S_\lambda(q,\lambda) &= \overline{\sum_k Q^k(q)\delta(\lambda-\lambda^k)},\\
    Q^k(q) &= \frac{1}{N}\left|\sum_j e^k_j\; \exp(i\, q x_j)\right|^2.
\end{align}
Here, we have assumed an excitation along the $x$-axis with $\vec{q} = q\,\vec{\hat{e}}_x$. Note that throughout all analyses we only allow discrete wavevectors $q_l = l\,2\pi/L$ with $l=\pm1,\pm 2,\ldots$, satisfying the periodic boundary conditions. 
The dynamic structure factor can be used to  extract the dispersion relation $\Omega(q)$ and the damping $\Gamma(q)$ by fitting $S(q,\omega)$ to a damped harmonic oscillator model~\cite{Ramos2022,Shintani2008}
\begin{align}\label{eq:dampedOsci}
    S(q,\omega)\propto \frac{\Omega^2(q)\Gamma(q)}{(\omega^2-\Omega^2(q))^2+\omega^2\Gamma^2(q)}
\end{align}
Another quantity, which is used to characterise the eigenmodes of the systems, is the participation ratio
\cite{Shimada2018,Mizuno2017,Wang2019-2,Hu2022}

\begin{align}
        P^k = \frac{1}{N} \frac{1}{\sum_i (e_i^k\, e_i^k)^2}.
\end{align}
The participation ratio $P^k = 1/N \ll 1$ is small for an ideal localized mode involving only one particle, while $P^k = 2/3$ for an ideal plane wave\add{~\cite{Ramos2022}, and $\overline{P^k} = 0.3$ if the eigenvectors are isotropically oriented \cite{Beltukov2011}}. 

\add{The level statistics $p(s)$ provides insight into the coupling of eigenmodes. It describes the distribution of the distances between neighboring eigenvalues. The levels are randomly distributed according to the Poisson distribution $p(s) = \exp(-s)$ in the case of localized modes. For delocalized states in the GOE ensemble, Wigner's surmise should hold: $p(s) = \frac{1}{2}\pi\,s\exp(-\pi\,s^2/4)$ \cite{Mehta2014}. To numerically evaluate $p(s)$, we first have to compute the normalized distance $s^k$ between neighboring eigenfrequencies. For this, we calculate the difference between successive eigenfrequencies $r^k = |\omega^k-\omega^{k+1}|$ and divide it by the mean distance between levels $\overline{r^k}$~\cite{Beltukov2013}, which gives
\begin{align}
    s^k = \frac{r^k}{\overline{r^k}}.
\end{align}
We can then calculate $p(s)$ in the vicinity of a specific eigenfrequency $\omega_0$ by computing a histogram of the distances $s^k$ in an interval of width $\Delta \omega$ around $\omega_0$. In the following $\Delta \omega$ is fixed to a value of $\Delta \omega = 0.1$.
}

\subsection{Excited wave analysis}
If the eigenvectors $\vec{e}^k$ of the system are known, we can also easily solve the equations of motion by
\begin{align}
    \phi_i(t) = \sum_k u^k(t)e_i^k,
\end{align}
where 
\begin{align}
    u^k(t) = \vec{e}^k\cdot \vec{\phi}(0) \cos(\omega^k t) + \vec{e}^k\cdot \dot{\vec{\phi}}(0)\frac{\sin(\omega^k t)}{\omega^k}.
\end{align}
The scalar product abbreviates the sum over particles, $\vec{e}^k\cdot \vec{a}=\sum_i e^k_i\, a_i $. We can excite a standing wave with wavenumber $q$  in our inherent structure by choosing for example $\phi_i(0) = 0$ and $\dot{\phi}_i(0) =~\sin(q x_i + \Phi)$ as initial conditions~\cite{Mizuno2018,Wang2019,Saitoh2021,Gelin2016}, with $\Phi = 0, \pi/2$. Equivalent to fitting the damped harmonic oscillator model to the dynamic structure factor in the frequency domain, we can calculate the correlation function
\begin{align}
    C(q,t) &= \overline{R(q,t)},\\
    R(q,t) &= \frac{\sum_i \dot{\phi}_i(0)\,\dot{\phi}_i(t)}{\sum_i \dot{\phi}_i(0)\, \dot{\phi}_i(0)}\, 
\end{align}
and fit it with 
\begin{align}
     C(q,t) = \exp(-\Gamma(q)\, t/2)\cos(\Omega(q)t).
\end{align}
Note that the average now also includes both phases $\Phi = 0, \pi/2$ for each set of inherent positions.
This again allows us to extract the dispersion relation $\Omega(q)$ and the attenuation $\Gamma(q)$. While the dynamic structure factor includes a binning process in its calculation, the correlation function does not, yielding better results for small wavevectors $q$.\\

We can rewrite the correlation function in terms of "hybridization" coefficients 
\begin{align}
    \xi^k(q)= \frac{\vec{e}^k\cdot \dot{\vec{\phi}}(0)}{\sqrt{\dot{\vec{\phi}}(0)\cdot\dot{\vec{\phi}}(0)}}    
\end{align}
by
\begin{align}
    R(q,t) &= \frac{\sum_k \vec{e}^k\cdot \dot{\vec{\phi}}(0) \cos(\omega^kt) \vec{e}^k \cdot \dot{\vec{\phi}}(0)}{\dot{\vec{\phi}}(0)\cdot\dot{\vec{\phi}}(0)}
    \\ &= \sum_k (\xi^k(q))^2 \cos(\omega^k t) ,
\end{align}
where $\sum_k (\xi^k(q))^2 = 1$ must hold.
\section{Numerical details}\label{sec:numerical}
We study systems with periodic boundary conditions. In our case this means that the periodic boundaries influence the calculation of the dynamical matrix $\Vec{M}$. A particle at the boundary of the simulation box also interacts with  the particles of the periodic copies of the simulation box. 
This enforces the selection of wavevectors $q_l$ introduced above.\\
In order to use sparse matrices (storing less then $N^2$ entries) for the large systems, we introduce a cut-off radius $\sigma$ at which the spring function $f(r)$ is truncated. We choose $\sigma = 4$. See below for a discussion of the accruing errors.\\
In order to obtain the randomly generated inherent structures we use the standard random number generator of \textit{matlab} which is the Mersenne Twister algorithm~\cite{Matsumoto1998}.\\
The main simulation parameter we will vary is the dimensionless density $n$. At a given number of particles $N$ the density determines the size of the periodic simulation box \textit{via} $L = (N/n)^{1/d}$. Hence, if we increase the number of particles $N$ we also increase the size of the simulation box $L$. At the same time we also get access to smaller wave vectors $q$.

We calculate the full set of eigenvalues for system sizes up to $N = 4\times 10^4$ particles and the smallest $2000$ eigenvalues for larger systems. This drastically reduces the computation time and especially the storage consumption. See below for a discussion of the accruing errors. 

For each density $n$ and system size $N$ we perform calculations in $250$ realizations of the random inherent positions and present the averages.  When calculating the 3D participation ratios, we tested an ensemble size of $5\times10^5$ simulations and found qualitatively the same result.

\section{Results and discussions}
First, we show the results obtained \textit{via} the normal mode analysis and afterwards we show the excited wave analysis. \add{Then we increase the ensemble size to analyse the structure of the eigenmodes.}

\subsection{Density of states}

\begin{figure}[H]
    \centering
    \begin{subfigure}[b]{0.45\textwidth}
    \includegraphics[width=0.99\textwidth]{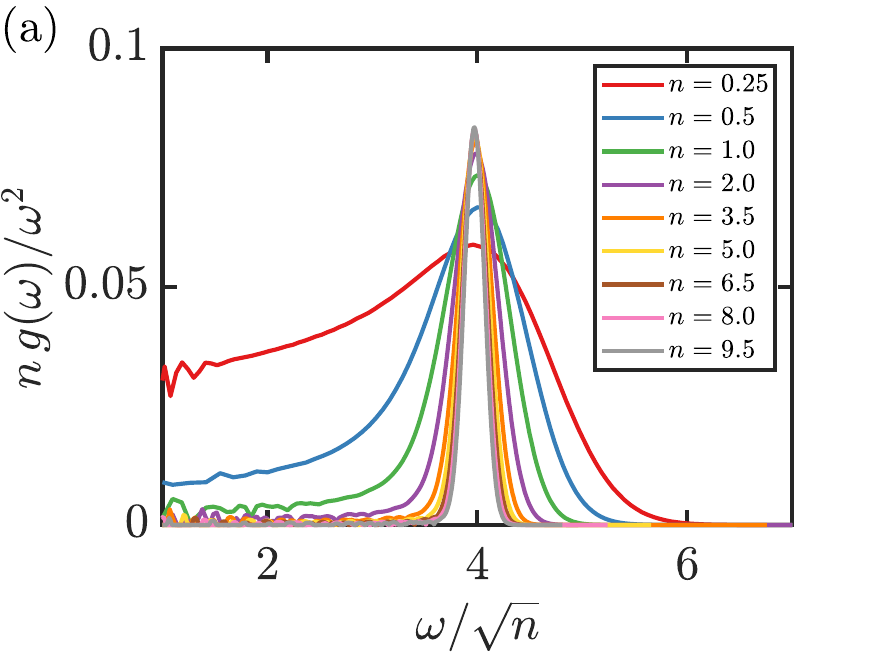}
    \end{subfigure}
    \begin{subfigure}[b]{0.45\textwidth}
    \includegraphics[width=0.99\textwidth]{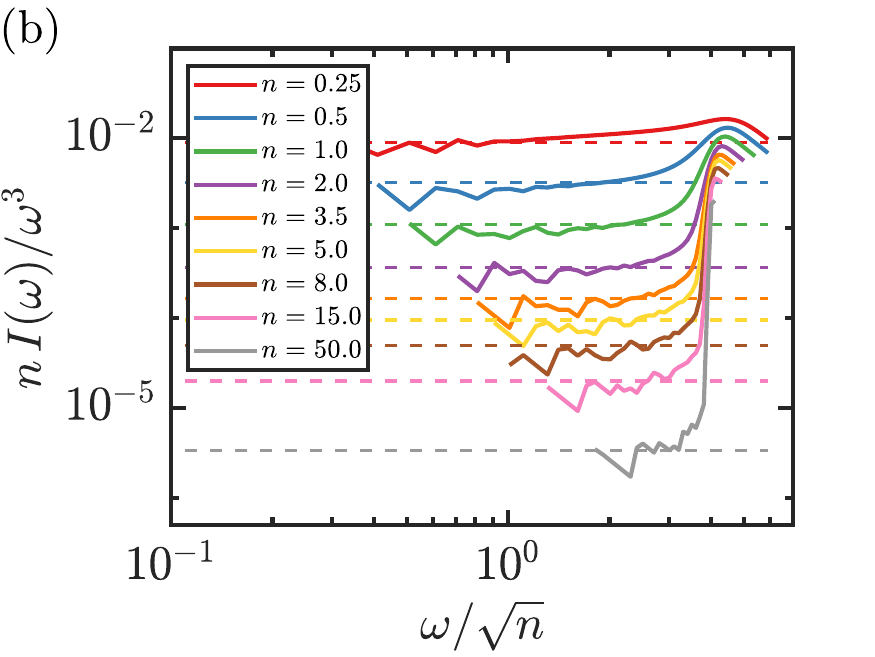}
    \end{subfigure}
    \caption{\textbf{(a)} Density of states $g(\omega)$ and \textbf{(b)} integrated density of states $I(\omega)$ divided by the Debye behaviour for different densities $n$ (see legend) and as function of the rescaled frequency. The dashed lines indicate the Debye levels $n A_\mathrm{D}/3$. \add{From top to bottom the density increases.}}
    \label{fig:dos}
\end{figure}

Fig.~\ref{fig:dos}(a) shows the density of states for systems of size $N=4\times 10^4$ for varying densities $n$. As is often done, 
we look at the  reduced vDOS, $g(\omega)/\omega^2$. This anticipates the Debye-law at low frequency,  $g_\mathrm{D}(\omega) = A_\mathrm{D}\, \omega^2 = \omega^2/\omega_\mathrm{D}^3$, where $A_\mathrm{D}$ is the Debye level and $\omega_\mathrm{D}$ the Debye frequency. \add{Both of these quantities are calculated here from the vDOS in the small $\omega$-limit}.
We can observe a Debye spectrum for $\omega\to 0$. 
It is expected because of the breaking of translational invariance in the solid inherent structures which leads to the existence of sound waves for small wavevector \cite{Grigera2001}. For lowering $n$, viz.~increasing average separation of the inherent positions, the elastic restoring forces become weaker and the Debye-frequency decreases. We observe an excess in the vDOS above the Debye-law for larger frequencies $\omega$, which we call the boson peak of the ERM. This interpretation will be discussed in Sect.~\ref{sec:conclusions}.
The location of the boson peak $\omega_\mathrm{BP}$ scales with $\sqrt{n}$. Fig.~\ref{fig:dos}(b) shows the integrated density of states $I(\omega)$ divided by the Debye behaviour for the same densities $n$. The dashed lines indicate the Debye levels $A_\mathrm{D}/3$, which can most directly be extracted from these rescaled $I(\omega)$ data.\\
\begin{figure}[H]
    \centering
    \begin{subfigure}[b]{0.45\textwidth}
    \includegraphics[width=0.99\textwidth]{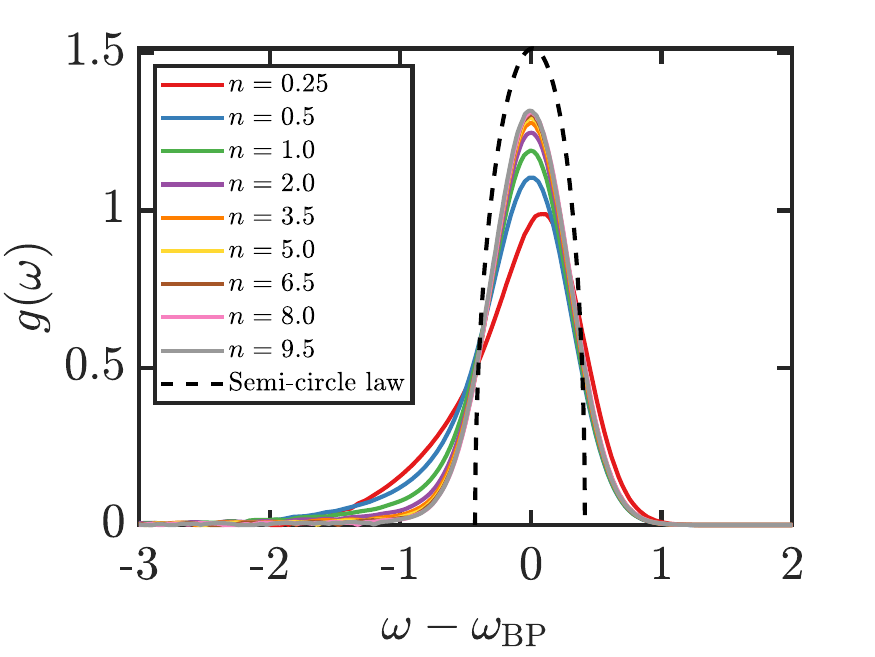}
    \end{subfigure}
    \caption{Density of states $g(\omega)$  for different densities $n$ (see legend) plotted as function of $\omega-\omega_\mathrm{BP}$. The dashed line shows the predicted Wigner semicircle law \cite{Vogel2023}. \add{The height of the peak increases for increasing density. }
    }
    \label{fig:dos2}
\end{figure}
Fig.~\ref{fig:dos2} shows the vDOS presented as function of $\omega-\omega_\mathrm{BP}$. In this plot the curves coincide quite well \cite{fuss}, highlighting that the majority of eigenfrequencies is accumulated in the boson peak. The overall shape of the vDOS is well described by Wigner's semicircle law, which would hold if the entries of $\bf M$ were independent and identically distributed zero-mean Gaussian entries \cite{Skipetrov2013}. The semicircle law is shown by the dashed lines in Fig.~\ref{fig:dos2}; note that for simplicity its normalization to unity is not adjusted to fit the data best. The semicircle is (in energy space) located around $\omega_\mathrm{BP}^2$ and has a radius of $R$. In frequency space the explicit form is given by 
\begin{align}
    g_\mathrm{wigner}(\omega) = \frac{4\,\omega}{\pi R^2}\sqrt{R^2-(\omega^2-\omega_\mathrm{BP}^2)^2}\;,
\end{align} 
\add{for $\omega_- \le \omega \le \omega_+$ with $\omega_\pm=\sqrt{\omega^2_{BP}\pm R}$.} \add{We find good agreement with the theory by taking the values $R = n\sqrt{2a}$ with $a = \hat{f}(0)/(\sqrt{8}n)$ from Ref.~\cite{Vogel2023}.}

Note that $g_\mathrm{wigner}(\omega_\mathrm{BP}) = 4\,\omega_\mathrm{BP}/(\pi R) = 4/\pi\, 2^{1/4} \propto n^0$\add{, for large $n$}. A shift of the boson peak frequency  is the dominant effect when changing the disorder via changing $n$. The plot hides the $n$-dependence of the vDOS for small frequencies, as  the  Debye-level is very low compared to the boson peak amplitude for the considered $n$.

\begin{figure}[H]
    \centering
    \begin{subfigure}[b]{0.45\textwidth}
        \includegraphics[width=1\textwidth]{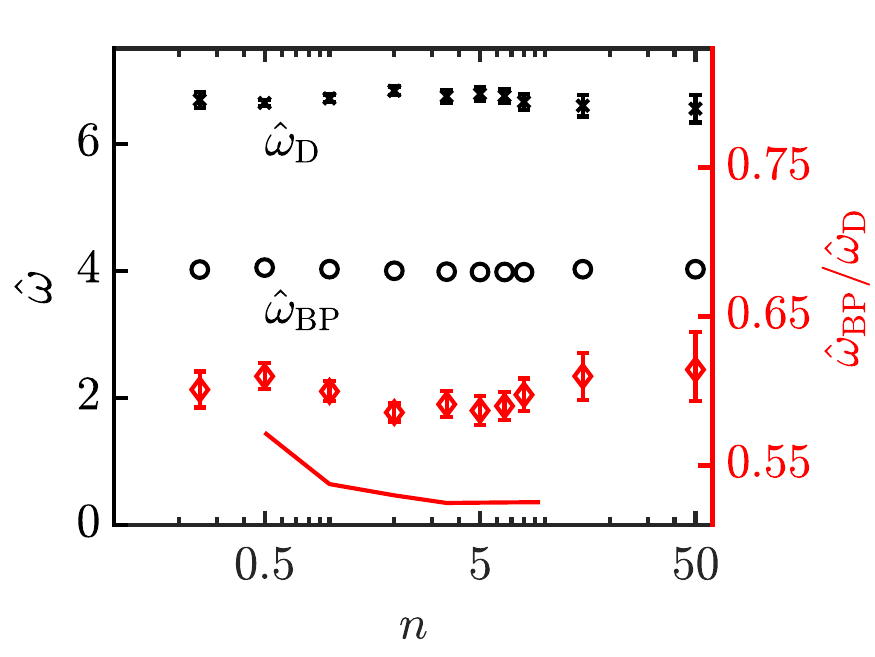}
    \end{subfigure}
    \caption{    Rescaled Debye frequency $\hat{\omega}_\mathrm{D} =\omega_\mathrm{D}\, n^{-5/6}$ (black circles) and boson peak frequency $\hat{\omega}_\mathrm{BP} =\omega_\mathrm{BP}\, n^{-1/2}$ (black crosses) as functions of the density $n$. The $\hat{\omega}_\mathrm{D}$ and $\hat{\omega}_\mathrm{BP}$ are $n$-independent in stable states. Additionally, the ratio $\hat{\omega}_\mathrm{BP}/\hat{\omega}_\mathrm{D}$ extracted from the simulation data (red diamonds\add{, right axis}) is shown and compared to theory (line) from Ref.~\protect\cite{Vogel2023}.}
    \label{fig:frequencies}
\end{figure}

From $g(\omega)$ and $I(\omega)$ we extract the density dependence of the relevant frequencies, $\omega_\mathrm{D}$ and $\omega_\mathrm{BP}$ shown in Fig.~\ref{fig:frequencies}. The uncertainty of $\omega_\mathrm{D}$ is calculated from the confidence interval of the fit of the Debye-level $A_\mathrm{D}$ and the uncertainty of $\omega_\mathrm{BP}$ is estimated by evaluating the frequencies of the bins left and right of the maximum of the vDOS. Since the uncertainties in $\omega_\mathrm{BP}$ are very small they are omitted throughout this work. \add{The energy scale of the boson peak arises from the pairwise interaction among all particles \cite{Vogel2023}. This explains the scaling $\omega_\mathrm{BP}\propto n^{1/2}$.} For high enough $n$, the boson peak position is the square-root of the mean value of the diagonal entries  $\omega_\mathrm{BP}^2 = \overline{M_{ii}}=n \hat{f}(0)$ 
\cite{Vogel2023}. 

Considering the amplitude of the Debye-law, $A_D=1/\omega_D^3$, we observe that $A_D\propto n^{-5/2}$ and thus $\omega_\mathrm{D}$ scales with $n^{5/6}$.  The increase of the Debye frequency with decreasing $n$ is a non-trivial effect of the increasing disorder, which will be explained based on the dispersion relations shown in Fig.~\ref{fig:dispRel} below. While the obtained boson peak positions $\omega_\mathrm{BP}$ perfectly match with the values from the theory of~\cite{Vogel2023}, we find small deviations in the Debye frequency $\omega_\mathrm{D}$ and therefore also in the ratios $\omega_\mathrm{BP}/\omega_\mathrm{D}$. Still, the scaling of $\omega_\mathrm{D}$ and $\omega_\mathrm{BP}/\omega_\mathrm{D}$ with the density $n$ matches well with the theory.

\subsection{Dynamic structure factor and dispersion relation}
\begin{figure}[H]
    \begin{subfigure}[b]{0.45\textwidth}
    \includegraphics[width=0.99\textwidth]{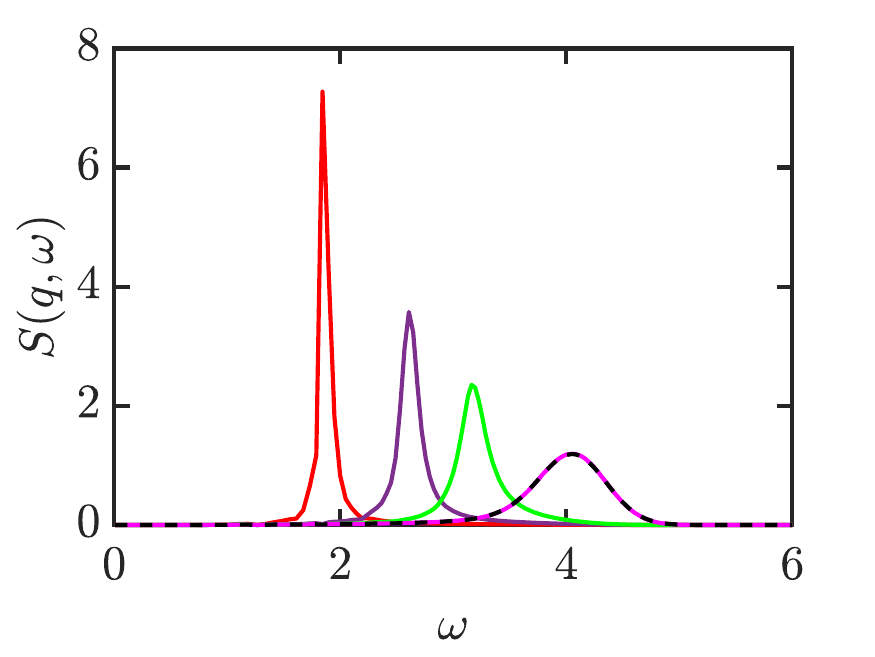}
    \end{subfigure}
    \caption{Dynamic structure factor $S(q,\omega)$ for $q_4, q_6,q_8$ and $q_{25}$ (marked in Fig.~\ref{fig:dispRel}; $q$ increases from left to right) at $n = 1.0$. The density of states \add{$g(\omega)$} is shown by the black dashed line, which agrees with $S(q_{25},\omega)$.
    }
    \label{fig:dyn}
\end{figure}

\begin{figure}[H]
    \begin{subfigure}[b]{0.45\textwidth}
    \includegraphics[width=0.99\textwidth]{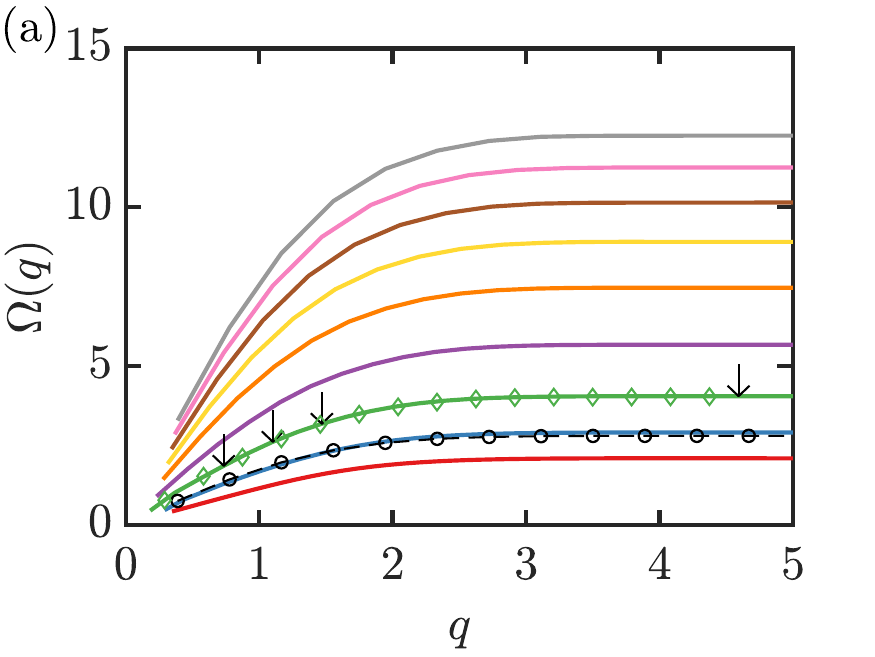}
    \end{subfigure}
    \begin{subfigure}[b]{0.45\textwidth}
    \includegraphics[width=0.99\textwidth]{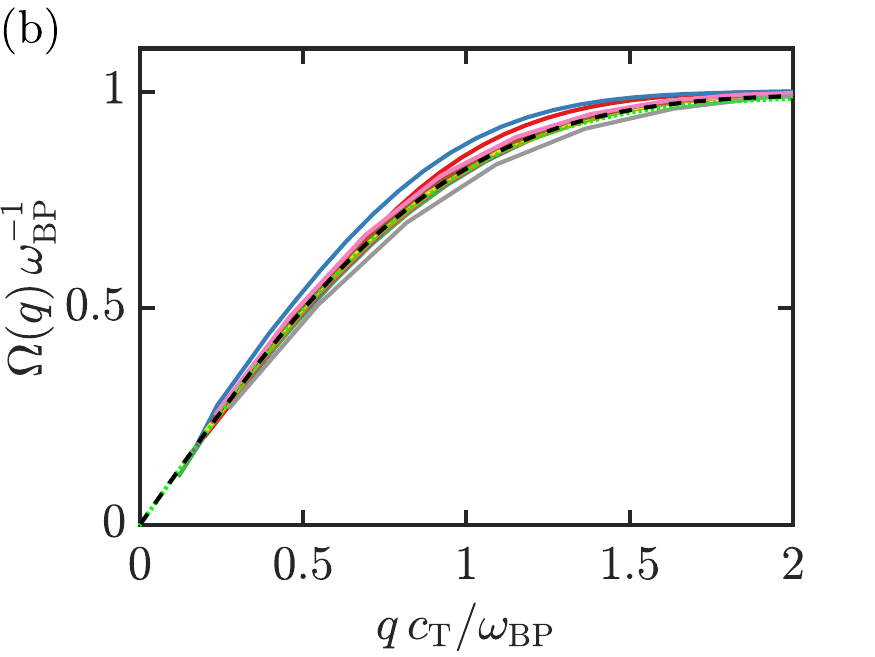}
    \end{subfigure}
    \caption{ \textbf{(a)} Dispersion relation $\Omega(q)$ for varying densities $n$ (see legend in Fig.~\ref{fig:dos2}). \add{The density increases from bottom to top.} The black arrows at $n = 1.0$ mark the $q$ values at which the dynamic structure factors in Fig.~\ref{fig:dyn} are shown. The black circles at $n = 0.5$ indicate the bare dispersion $\sqrt{\epsilon_0(q) n} = \sqrt{(\hat{f}(0)-\hat{f}(q))n}$ and the green diamonds show the dispersion relation obtained from the excited wave analysis.  \textbf{(b)} Rescaled dispersion relation $\Omega(q)\,\omega_\mathrm{BP}^{-1}$ in dependency of the rescaled wave vector $q\,c_\mathrm{T}/\omega_\mathrm{BP}$. Additionally, the prediction from Ref.~\cite{Vogel2023} (black dashed line) and from Ref.~\cite{Grigera2001} (green \add{dotted} line) both at $n=1.0$ are shown.}
    \label{fig:dispRel}
\end{figure}

Fig.~\ref{fig:dyn} shows the dynamic structure factors $S(q,\omega)$ at $n = 1.0$ for four different $q$ values. The structure factors are characterised by a pronounced peak which shifts to higher frequencies $\omega$ with increasing wave vector $q$. At the same time the peak broadens and its height decreases. \add{In the limit of $q\to \infty$, $S(q,\omega)$ approaches the vDOS $g(\omega)$ shown by the dashed line in Fig.~\ref{fig:dyn}  \cite{Martin-Mayor2001}.}\\
We use Eq.~\eqref{eq:dampedOsci} to extract the dispersion relation $\Omega(q)$ from the structure factor.  The results are shown in Fig.~\ref{fig:dispRel}(a). We observe a linear dispersion relation $\Omega(q)= c_\mathrm{T} q$ for small wave vectors and that $\Omega(q)\to\omega_{BP}$ saturates for $q\to \infty$. Here, $c_\mathrm{T}$ denotes the speed of sound, it inherits the scaling $c_T\propto \sqrt n$ from the dispersion relation. For increasing density $n$ the slope of the dispersion relation, and therefore, the speed of sound $c_\mathrm{T}$ gets larger; the system gets stiffer with the average separation of particles getting smaller. \add{The scaling $\Omega(q)\propto \sqrt{n}$ arises from the pair-wise interactions \cite{Vogel2023}.} The dispersion relation obtained \textit{via} the excited wave analysis is shown exemplarily for $n=1.0$ by the purple circles and agrees perfectly with the dispersion relation from the structure factor.\\
We compare the dispersion relation with the bare dispersion, $\Omega_0(q) = \sqrt{\epsilon_0(q) n} = \sqrt{(\hat{f}(0)-\hat{f}(q))n}$, at \add{a small} density $n = 0.5$.  The largest deviations from the bare dispersion should be observed at the biggest disorder, i.e. the smallest density. Yet, at \add{this large} disorder we still observe a good agreement between the bare dispersion and $\Omega(q)$.  Comparing with the self-consistent theories \cite{Ciliberti2003,Vogel2023}, one notices that both overestimate the effects of disorder; see Fig.~1 in Ref.~\onlinecite{Vogel2023}, where dressed and bare dispersion relations are shown. Both predict a stronger change of $\Omega(q)$ than observed, yet the more elaborate resummation including non-planar diagrams lies closer to the data. \\
In Fig.~\ref{fig:dispRel}(b) we show the rescaled dispersion relation $\Omega(q)\,\omega_\mathrm{BP}^{-1}$ as function of the rescaled wave vector $q/q_{BP}$ with $q_{BP}=\omega_\mathrm{BP}/\,c_\mathrm{T}$. The rescaled data collapse very well indicating that the boson peak frequency $\omega_\mathrm{BP}$ is the relevant frequency for the rescaling. Using the Debye-frequency does not lead to a comparable rescaling  (not shown).

The scaling of the sound velocity with density, $c_T\propto \sqrt n$, also explains the scaling of the Debye frequency, as $A_D \approx 1/( 2 n\pi^2 c_T^3)$ approximately  holds for high enough densities \cite{Martin-Mayor2001,  Vogel2023}.  The Debye amplitude $A_D$ grows for increasing disorder because the sound velocity softens. Additionally, $A_D$ is inversely proportional to the number of degrees of freedom, which become fewer with lowering $n$ at fixed volume. Both effects together cause the dependence $\omega_D\propto n^{5/6}$ in $d=3$ which is  appreciably stronger than the dependence of the boson peak frequency on disorder.

\subsection{Sound damping}

\begin{figure}[H]
    \centering
    \begin{subfigure}[b]{0.45\textwidth}
    \includegraphics[width=0.99\textwidth]{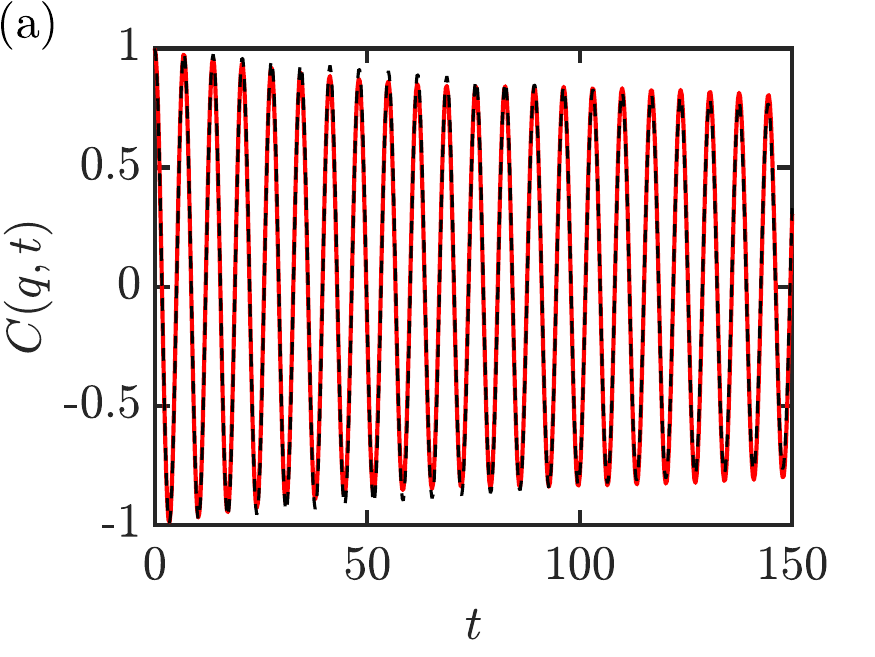}
    \end{subfigure}
    \begin{subfigure}[b]{0.45\textwidth}
    \includegraphics[width=0.99\textwidth]{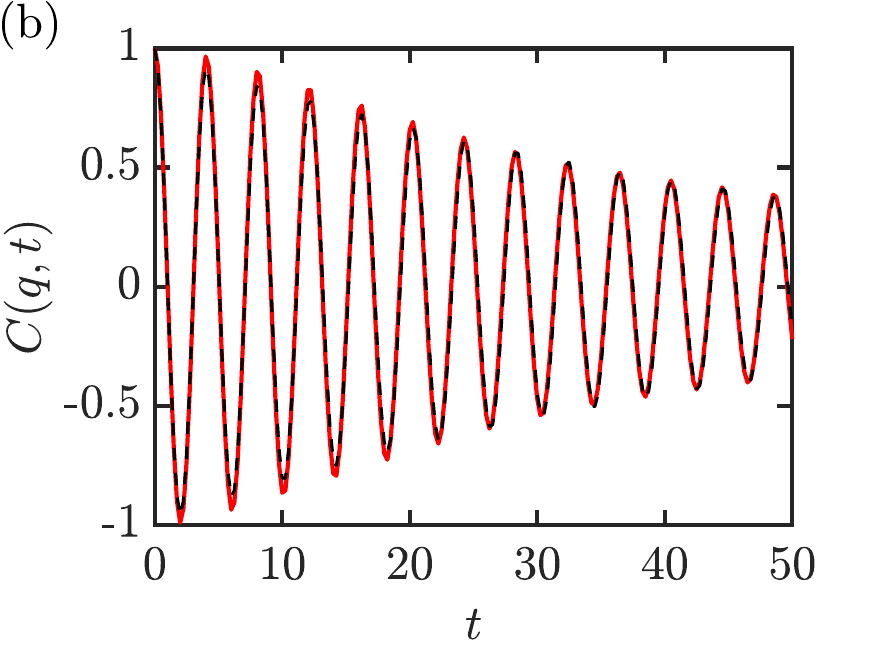}
    \end{subfigure}
    \caption{Velocity correlation function $C(q,t)$ of $N=4\times10^5$ systems at $n = 1.0$ for $q_4$ \textbf{(a)} and $q_7$ \textbf{(b)}. The black dashed lines correspond to a fitted damped oscillation. 
    }
    \label{fig:correlation}
\end{figure}

\begin{figure}[H]
    \centering
    \begin{subfigure}[b]{0.45\textwidth}
    \includegraphics[width=0.99\textwidth]{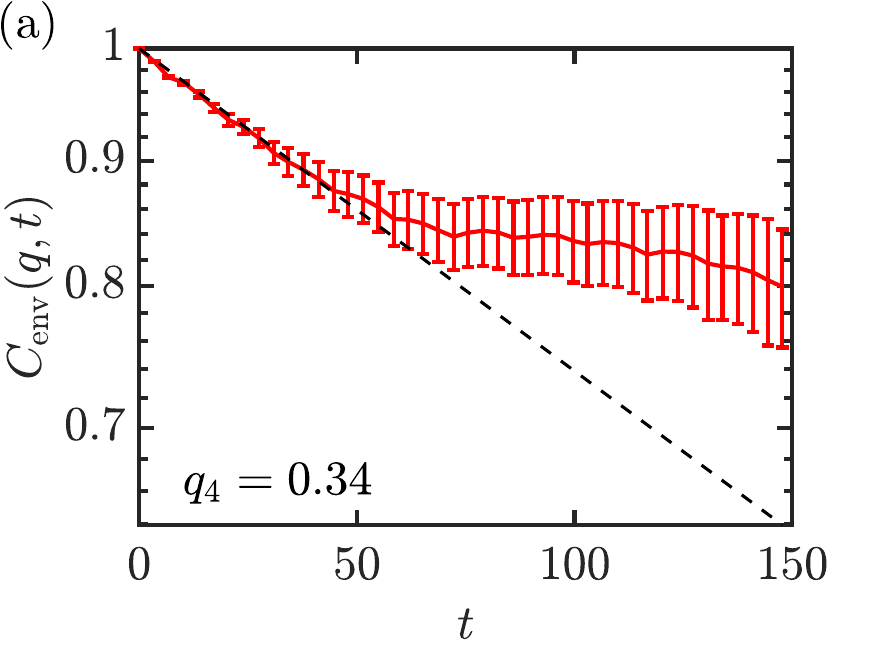}
    \end{subfigure}
    \begin{subfigure}[b]{0.45\textwidth}
    \includegraphics[width=0.99\textwidth]{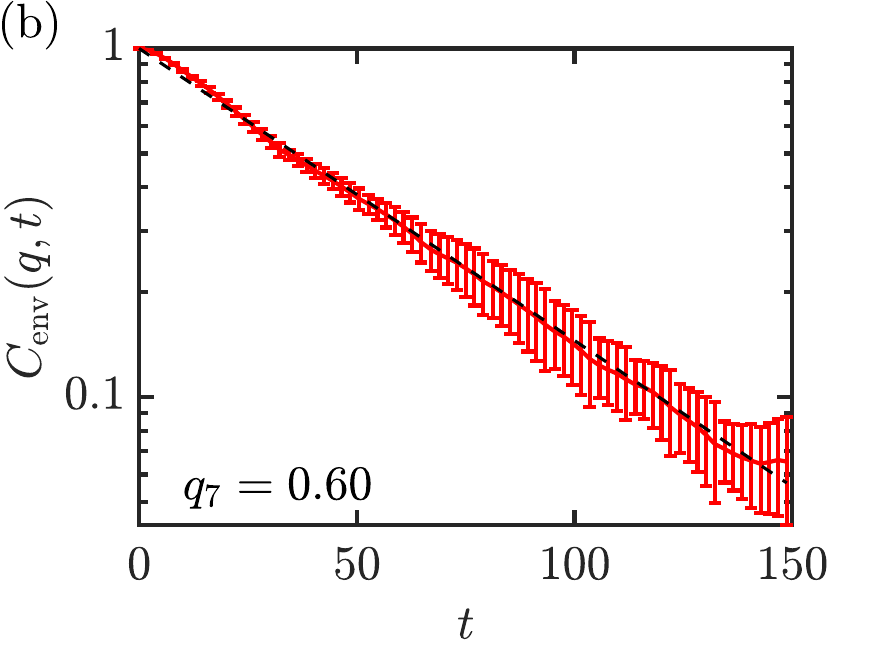}
    \end{subfigure}
    \begin{subfigure}[b]{0.45\textwidth}
    \includegraphics[width=0.99\textwidth]{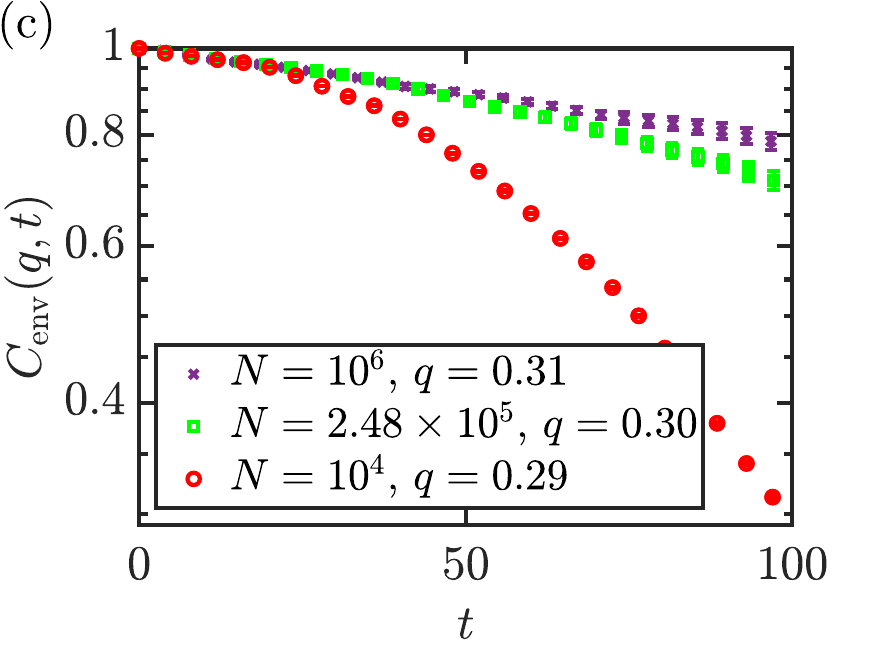}
    \end{subfigure}
    \caption{ Envelope of the velocity correlation functions shown in Fig.\ref{fig:correlation} for $q_4$ \textbf{(a)} and $q_7$ at $n = 1.0$ \textbf{(b)}. An exponential decay is fitted and is drawn by the black dashed line. \textbf{(c)} Envelope $C_\mathrm{env}(q,t)$ of three different system sizes at a similar wavevector.    }
    \label{fig:evelope}
\end{figure}

\begin{figure}[H]
    \centering
    \begin{subfigure}[b]{0.45\textwidth}
    \includegraphics[width=0.99\textwidth]{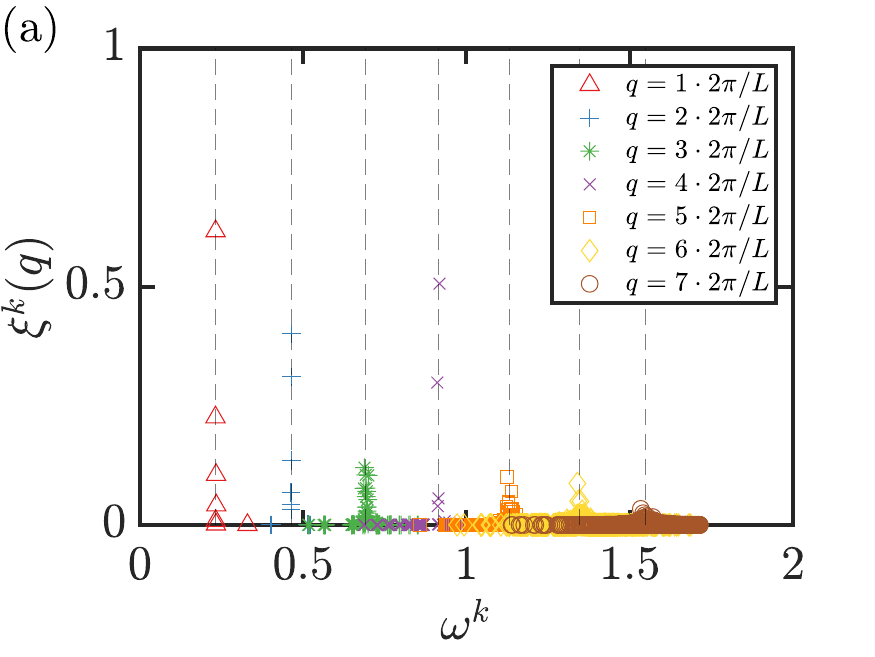}
    \end{subfigure}
    \begin{subfigure}[b]{0.45\textwidth}
    \includegraphics[width=0.99\textwidth]{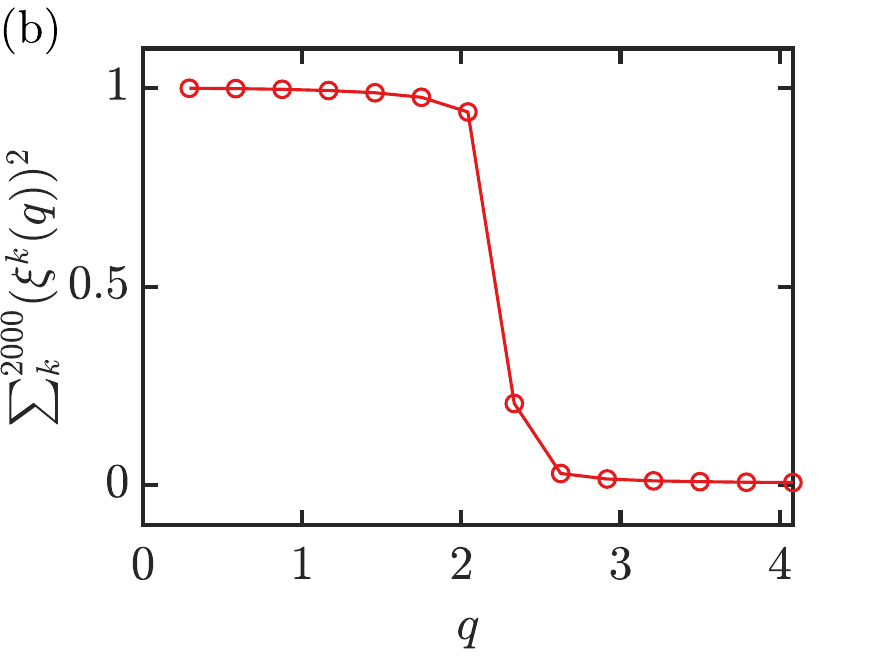}
    \end{subfigure}
    \caption{\textbf{(a)} Hybdridization coefficients for the $7$ smallest values of $q$ in a system at $n=1.0$ and for size $N = 4\times10^5$.  The larger $q$ becomes, the wider the range becomes over which non-vanishing hybridization coefficients are spread. The corresponding values of $\Omega(q)$ are indicated by the dashed grey lines. \textbf{(b)} The sum over the smallest $2000$ hybridization coefficients in this system drastically drops as soon as $q$ exceeds $q_7$. }
    \label{fig:hybrdization}
\end{figure}

\begin{figure}[H]
\centering    \includegraphics[width=0.5\textwidth]{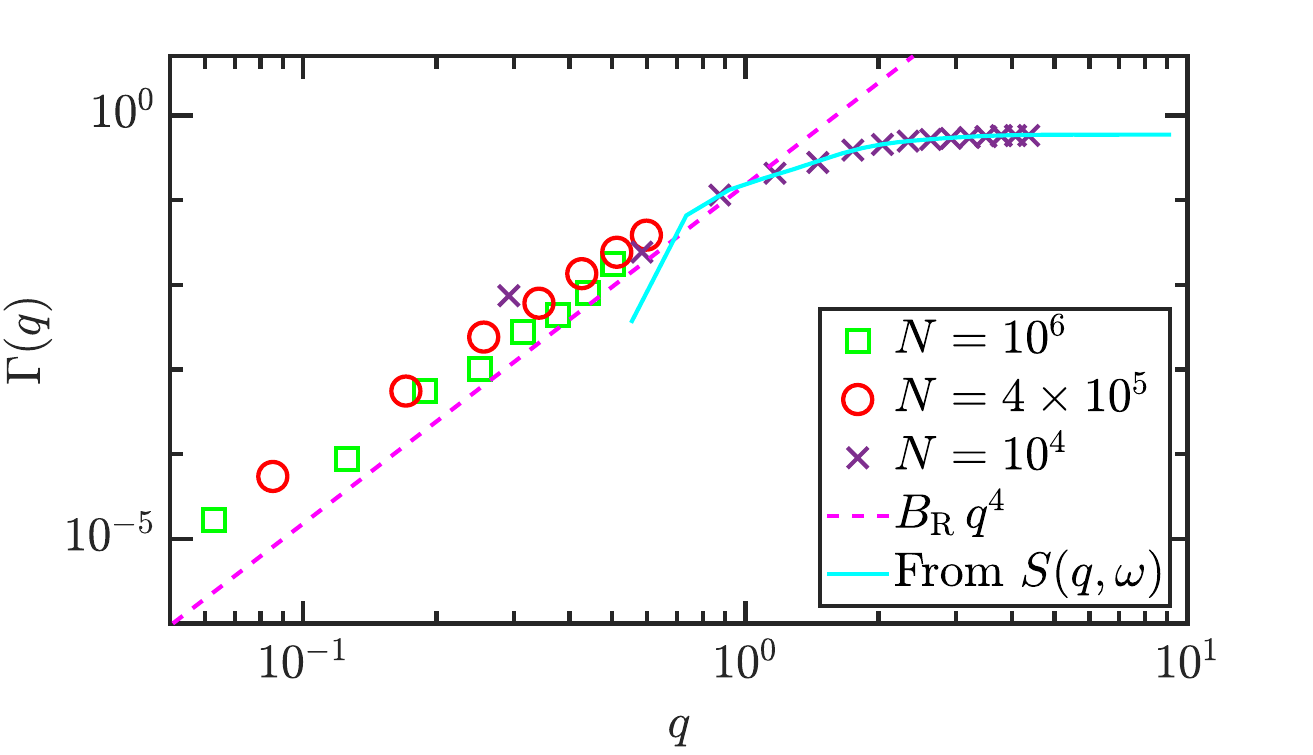}
    \caption{Sound attenuation $\Gamma(q)$ in dependency of the wave vector $q$ at $n = 1.0$. The data obtained from different systems sizes is combined. Note that the small systems are used to obtain the large $q$ behaviour, while the large systems are used to obtain the small $q$ behaviour. The \add{dashed} magenta  line corresponds to Rayleigh damping  $\Gamma(q) = B_\mathrm{R}q^4$ where the strength $B_\mathrm{R}$ is taken form the theory \cite{Vogel2023}.
    The blue line shows the attenuation as calculated from the dynamic structure factor $S(q,\omega)$.}
    \label{fig:damping}
\end{figure}

\add{In order to study the dissipation, we} turn to the excited wave analysis. Fig.~\ref{fig:correlation} shows the velocity correlation function $C(q,t)$ for $q_4$ and $q_7$. One can estimate the uncertainties of $C(q,t)$ by the standard deviation. For better visibility we do not shown them in Fig.~\ref{fig:correlation} because they are smaller than the symbols.\\
Clearly, a damped oscillation can be observed where the damping and the frequency becomes larger for larger $q$. For some $q$ there are beats with large magnitude for larger times $t$ which can be somewhat eliminated by averaging over different inherent structures. However, we get a more reliable result for the damping if we fit $\exp(-\Gamma(q)t)$ to the envelope $C_\mathrm{env}(q,t)$ of the correlation function\add{~\cite{Wang2019,Moriel2019}}.
Fig.~\ref{fig:evelope}(a) and Fig.\ref{fig:evelope}(b) show the envelope $C_\mathrm{env}(q,t)$ for the wavevectors $q_4$ and $q_7$ at $N = 4\times10^5$. Here, the uncertainties are those of $C(q,t)$ evaluated at the respective maxima and minima. The initial decay of the envelope is exponential (a fit is shown by the dashed black lines) but large deviations from the exponential decay are visible. This effect was already observed and discussed in~\cite{Wang2019,Bouchbinder2018} They argued that this is a finite size effect and that the deviations start at a system size dependent time and are stronger for smaller wavevector. Our results confirm this observation as can be seen in Fig.~\ref{fig:evelope}(c) where the envelope of the correlation function for three different systems sizes at a similar wavevector $q$ are shown. The uncertainty of the damping $\Gamma(q)$ therefore is due to the uncertainty of the range in which the envelope can by fitted by an exponential. 
So far, we neglected the fact that only the $2000$ smallest eigenvectors are calculated for systems with $N>4\times10^4$. We will now argue that the $2000$ smallest eigenvectors are sufficient to capture the small $q$ behaviour of the velocity correlation function. For this we look at the hybridization coefficients shown in Fig.~\ref{fig:hybrdization}.\\
As can be seen, for small values $q$ we get very large hybridization coefficients only for a narrow band of frequencies. As $q$ increases, a wider range of frequencies is involved in the response of the system to a standing wave. At a certain value of $q$ the frequencies with large hybridization coefficients start to lie outside of the smallest $2000$ eigenvalues. At this point the approximation of using only the $2000$ smallest eigenvectors necessarily fails. In Fig.~\ref{fig:hybrdization}(b) we show the sum over the hybridization coefficients for different values of $q$. As long as this sum is $1$ our approximation for $C(q,t)$ is valid. If the sum distinctively differs from $1$ we have not included enough frequencies and the approximation fails. For the shown system this is the case at $q> q_7$.\\
Finally, we show the obtained damping $\Gamma(q)$ in Fig.~\ref{fig:damping}. We observe that the attenuation becomes larger for increasing wave vectors $q$. The damping saturates for large $q$. For very small $q$ we observe a weak quartic (Rayleigh) damping which can be fitted by $\Gamma(q\to0)\to B_rq^4$. The theoretically predicted prefactor $B_R$ \cite{Vogel2023} lies within a factor of two relative to the exact prefactor. The theory prediction  $B_rq^4$ is included in Fig.~\ref{fig:damping}. In the high density regime holds $B_R \approx \frac{7}{48 \pi} \frac{\omega_{BP}^4}{n c_T^3}$ around the sound pole.

\begin{figure}[H]
    \centering
    \begin{subfigure}[b]{0.5\textwidth}
     \includegraphics[width=1\textwidth]{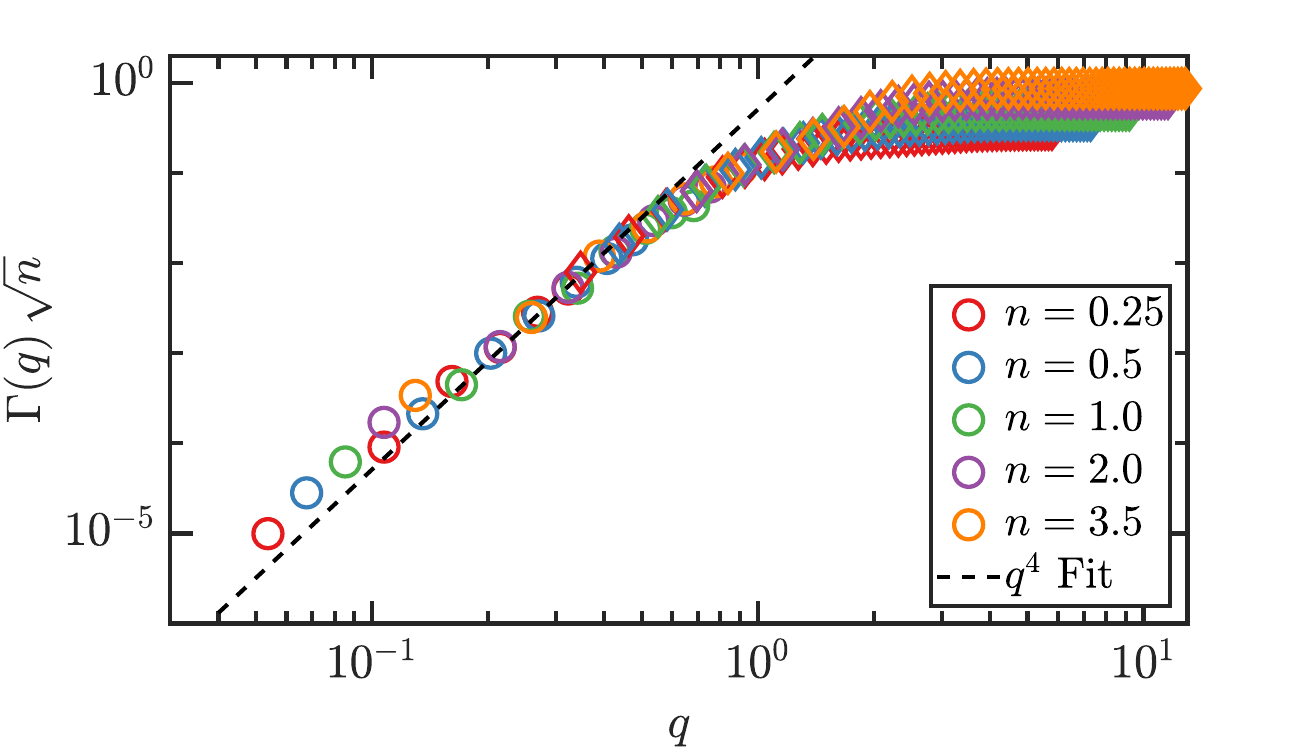}
    \end{subfigure}    
    \caption{\add{Rescaled sound attenuation $\Gamma(q)\,\sqrt{n}$ for different densities $n$. The colored squares depict the results at system size $N = 4\times 10^5$ and the colored circles the results at $N = 4\times 10^4$. The dashed black line depicts a $q^4$ fit.}}
    \label{fig:dampingRescaled}
\end{figure}

\add{The relation $B_R\propto \omega_{BP}^4/(nc_T^3)\propto n^{-1/2}$ also predicts the density dependence of the sound attenuation. 
Fig.~\ref{fig:dampingRescaled} shows the rescaled sound attenuation $\Gamma(q)\,\sqrt{n}$ obtained from the excited wave analysis for the system sizes $N = 4\times 10^4$ and $N = 4\times 10^5$ at increasing densities. The rescaling with $\sqrt{n}$ results in a rather good collapse to a universal quartic scaling law in the small wavenumber regime. The damping saturates at large wavenumbers to a density dependent value that increases with decreasing density (Note that this dependency is hidden in the rescaled plot)} 

\subsection{Characterization of eigenmodes}

\begin{figure}[H]
    \centering
     \begin{subfigure}[b]{0.45\textwidth}
     \includegraphics[width=1\textwidth]{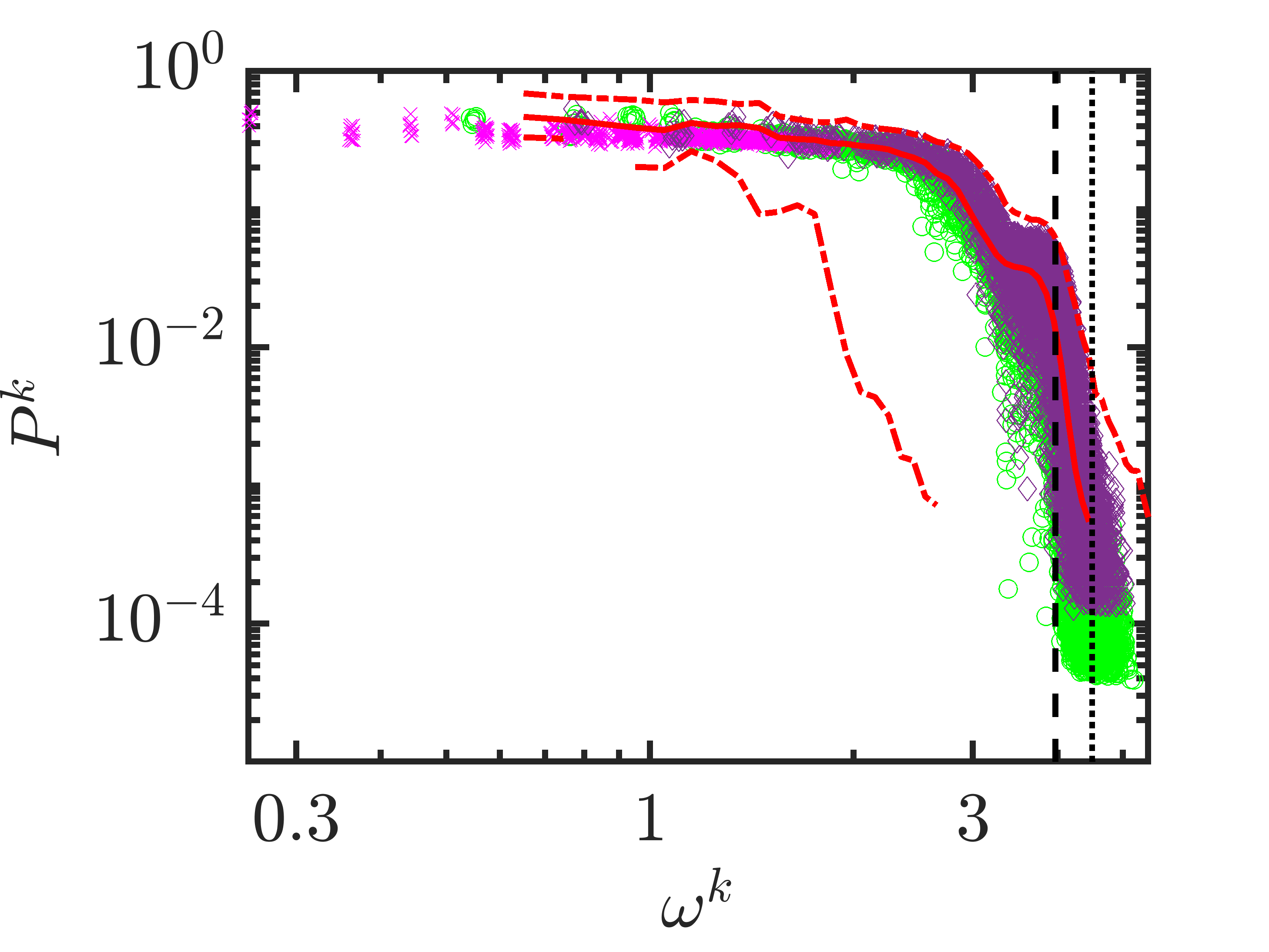}
    \end{subfigure}
    \caption{Participation ratio $P^k$ \add{at $n=1.0$} in dependency of the eigenfrequencies $\omega^k$ for system sizes $N = 10^4$ (purple \add{diamonds}), $N=3\times 10^4$ (green \add{circles}) and $N = 3\times 10^5$ (cyan \add{crosses}). Note that 
    in the last case,
    only the 2000 lowest participation ratios can be calculated. The black dashed line is located at the boson peak frequency \add{and the dotted line at $\omega_0 = 4.5$.} The red lines show the average, maximum and minimum participation ratio of an large ensemble of $N=10^4$ systems.}
    \label{fig:participationRatio}
\end{figure}

\add{After the study of the dynamic structure factor, we turn to the characterization of the eigenmodes. }
Fig.~\ref{fig:participationRatio} shows the participation ratios $P^k$ for a single realisation of the inherent positions at different system sizes $N$ at $n=1.0$. We can clearly see frequencies with large participation ratios of magnitude close to the \add{one of} ideal plane waves at small frequencies $\omega^k$. These frequencies have distinct gaps between them and belong to the phonon bands which have discrete frequencies due to the finite size of our system. The first participation ratios are therefore located at $\omega = c_\mathrm{T}2\pi/L$.\\
With increasing $\omega$, the participation ratios get smaller and at the boson peak they distinctively drop. Above the boson peak we observe localized modes. The behaviour is qualitatively the same for different system sizes. We have run additional simulations for an increased ensemble size of $5\times 10^5$ systems each with $N=10^4$. We calculate the average, minimum and maximum participation ratio in distinct frequency bins. The maximum and minimum participation ratio per bin is shown in Fig.~\ref{fig:participationRatio} by the red lines dashed dotted lines and the average participation ratio by the full line\add{; let $P(\omega)=\overline{P^k}$ denote this disorder average}. The large ensemble confirms the overall shape of the participation ratio distribution. Contrary to the interpretation in \cite{Vogel2023}, we do not find (quasi-) localised modes \cite{Lerner2021}. This may be a consequence of studying scalar excitations
\cite{Lerner2021, tanguy2023vibrations}.

\add{In order to further characterize the eigenmodes we analyze the level statistics $p(s)$. It describes the distribution of the distances between neighboring eigenvalues, and, in the case of localized modes, leads to a Poisson distribution $p(s) = \exp(-s)$, while for delocalized states in the GOE random matrix ensemble, it leads to Wigner's surmise: $p(s) = \frac{1}{2}\pi\,s\exp(-\pi\,s^2/4)$ \cite{Mehta2014}. 
}
\begin{figure}[H]
    \centering
    \begin{subfigure}[b]{0.45\textwidth}
     \includegraphics[width=1\textwidth]{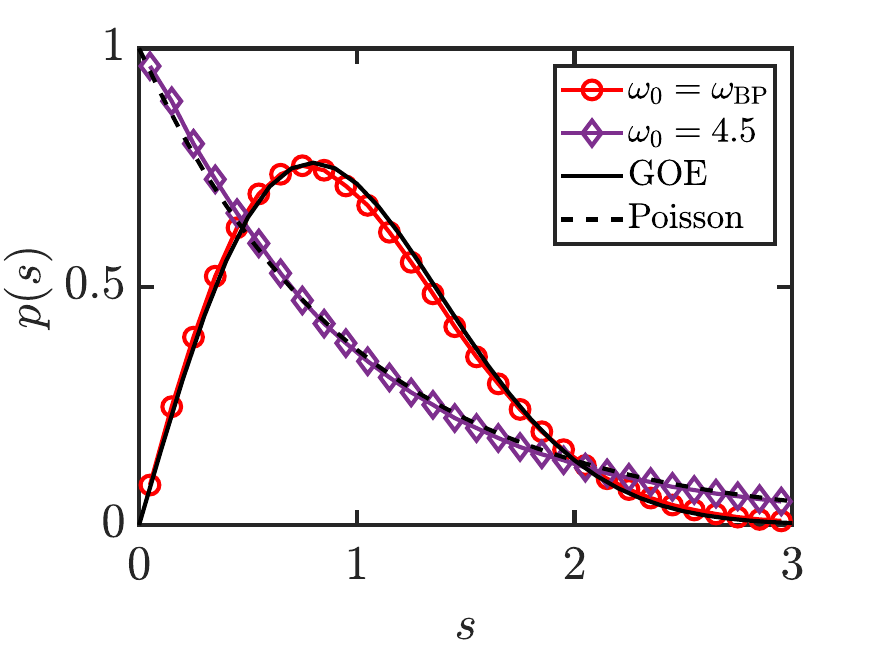}
    \end{subfigure}    

    \caption{\add{Level statistics $p(s)$ at $n=1.0$ for a frequency interval of width $\Delta \omega$ centred around $\omega_0 = \omega_\mathrm{BP}$ and $\omega_0 = 4.5=1.14\,\omega_{BP}$} and a comparison to the  GOE and Poisson statistics. \add{The locations of the frequencies $\omega_0$ are indicated in Fig.~\ref{fig:participationRatio} by the vertical dashed and dotted lines.}}
    \label{fig:levelStatistics}
\end{figure}
\add{Fig.~\ref{fig:levelStatistics} shows $p(s)$ around the Boson peak frequency $\omega_\mathrm{BP}$ and above it.
The level statistics nicely follows the GOE statistics around the Boson peak, while it decays according to the Poisson statistics at the higher frequency. As already suggested by the participation ratios, the eigenmodes are extended and of random matrix type in the frequency range of the Boson-peak \cite{Schirmacher1998} and localised above it.}\\

\begin{figure}[H]
    \centering
    \begin{subfigure}[b]{0.45\textwidth}
     \includegraphics[width=1\textwidth]{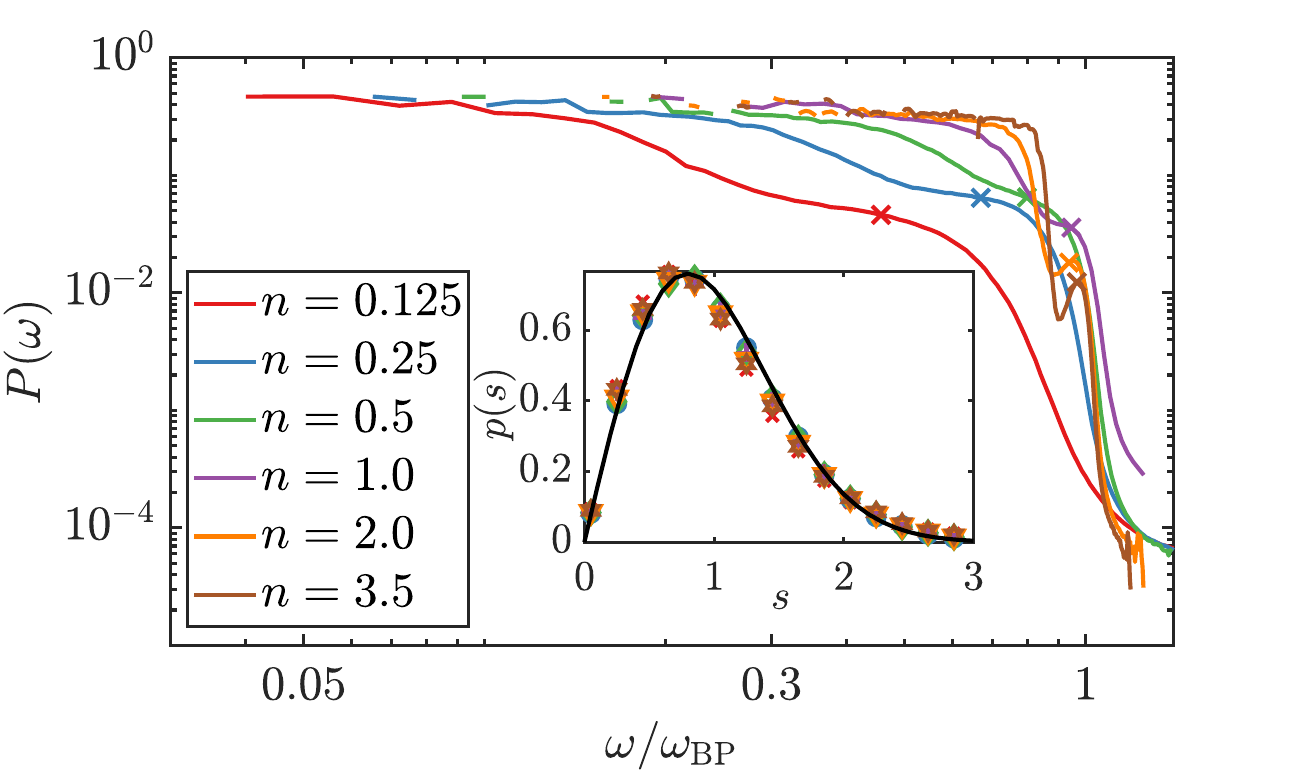}
    \end{subfigure}    
    \caption{\add{Mean participation ratio $P(\omega)$ as function of the rescaled frequency $\omega/\omega_\mathrm{BP}$ for a range of densities. The inset shows the level statistics $p(s)$ for frequencies that are marked by the coloured crosses. The GOE level statistics is shown by the black line.} 
    }
    \label{fig:ParticipationMean_LevelSpacing}
\end{figure}

\add{In Fig.~\ref{fig:ParticipationMean_LevelSpacing} we show the density dependence of the mean participation ratio $P(\omega)$ for  an extended $n$-range reaching rather low. For all densities shown, the Boson peak  marks a transition; above it all modes are localized. While for large densities there is a sharp transition to localized modes at exactly $\omega_\mathrm{BP}$, for decreasing densities the frequency above which the majority of the modes are localized shifts below the Boson peak. Below it, we observe a plateau region in $P(\omega)$ in a frequency window which depends on the density $n$; its lower edge  shifts to smaller frequencies for smaller density, viz.~for larger disorder, and its width increases.\\
By calculating the level statistics for modes in these plateau regions, we observe that they follow the GOE statistics. For arbitrarily chosen frequencies in the plateau regions, marked by the coloured crosses in Fig.~\ref{fig:ParticipationMean_LevelSpacing}, the corresponding GOE level statistics are shown in the inset of the figure. The increasing width of the plateaus for smaller densities indicates that the GOE modes gain in importance for small $n$. The number of oscillators equals $N=4\times 10^4$ in Fig.~\ref{fig:ParticipationMean_LevelSpacing}, but we observed no system-size dependence comparing with other $N$.}

\begin{figure}[H]
    \centering
    \begin{subfigure}[b]{0.45\textwidth}
     \includegraphics[width=1\textwidth]{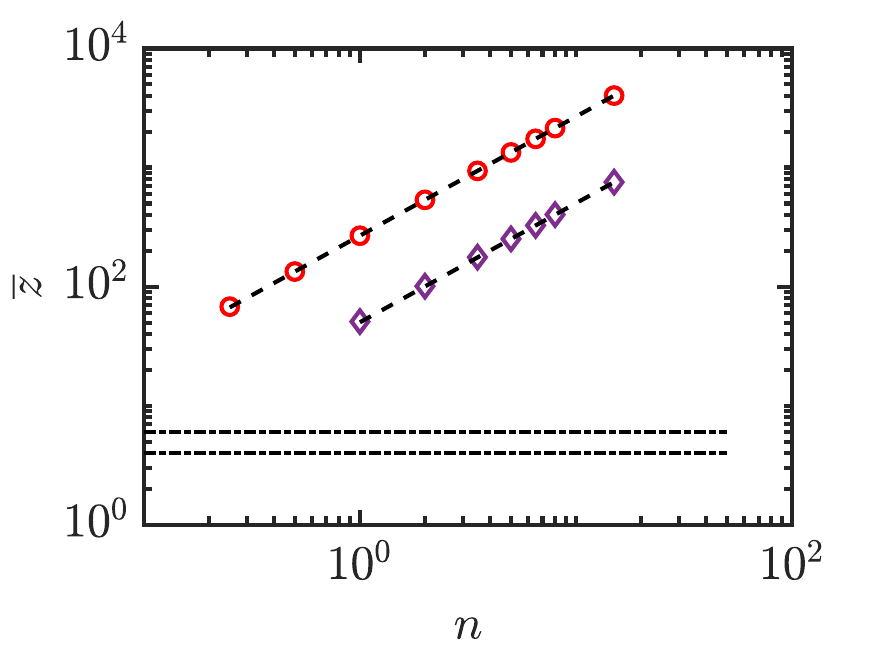}
    \end{subfigure}    
    \caption{Contact number $\overline{z}$ in dependency of the density $n$ for the 3D (red circles) and 2D (purple diamonds) systems. The dashed lines show the expected contact numbers and the dashed dotted lines are located at the isostatic stability criterion $\overline{z}_c = 2\,d$.}
    \label{fig:contactNumber}
\end{figure}

Fig.~\ref{fig:contactNumber} shows the contact number $\overline{z} = M/N -1$ where $M$ is the number of nonzero entries of $\Vec{M}$. Recall from Sect.~\ref{sec:numerical}, that a finite cut-off $\sigma$ was chosen in order to speed-up the matrix diagonalizations. Thus the contact number, which would be infinite for the Gaussian spring function in Eq.~\eqref{eq:springFunction}, becomes finite.
\begin{figure*}
    \centering
    \begin{subfigure}[b]{0.45\textwidth}
    \includegraphics[width=0.99\textwidth]{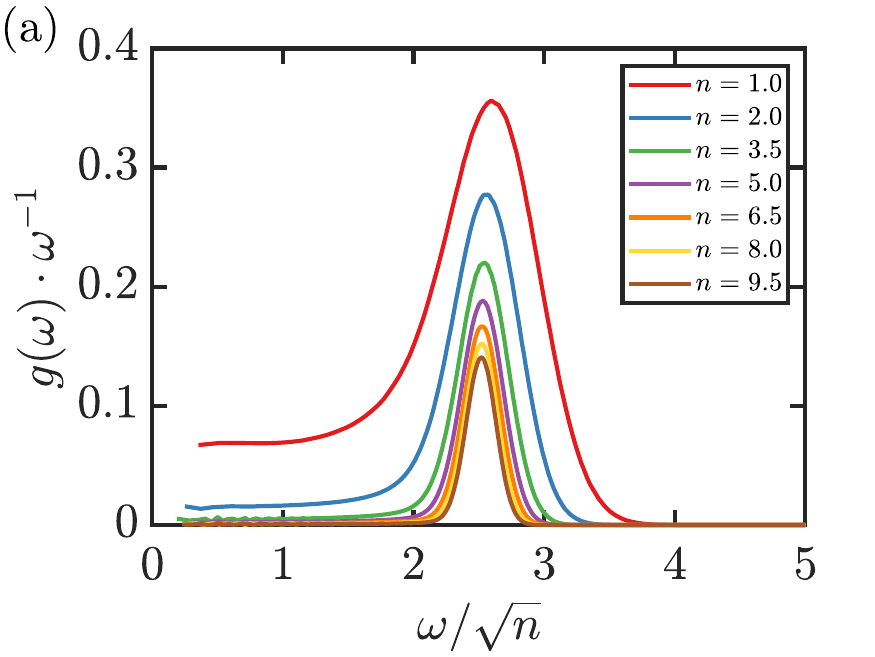}
    \end{subfigure}
    \begin{subfigure}[b]{0.45\textwidth}
    \includegraphics[width=0.99\textwidth]{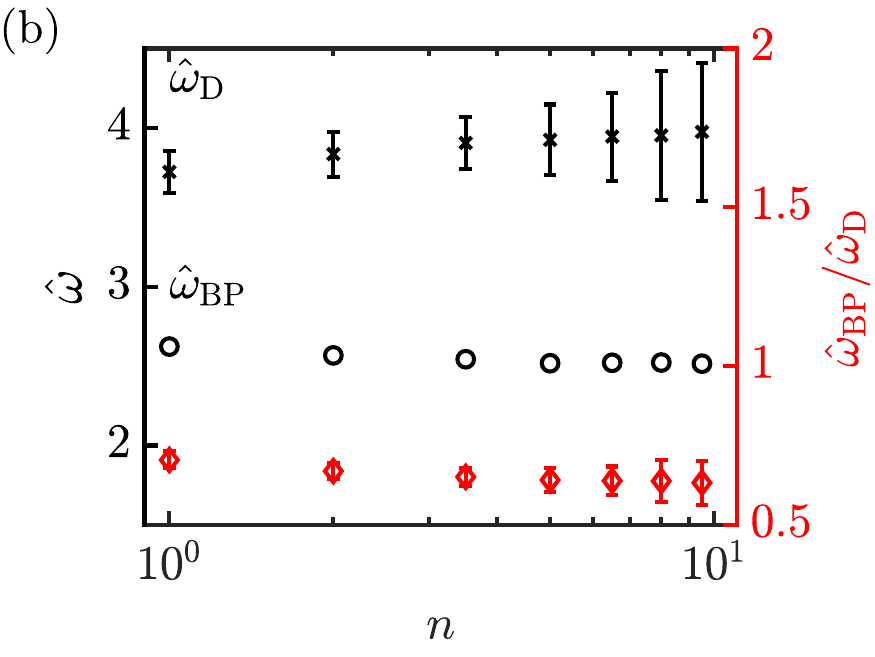}
    \end{subfigure}
    \begin{subfigure}[b]{0.45\textwidth}
    \includegraphics[width=0.99\textwidth]{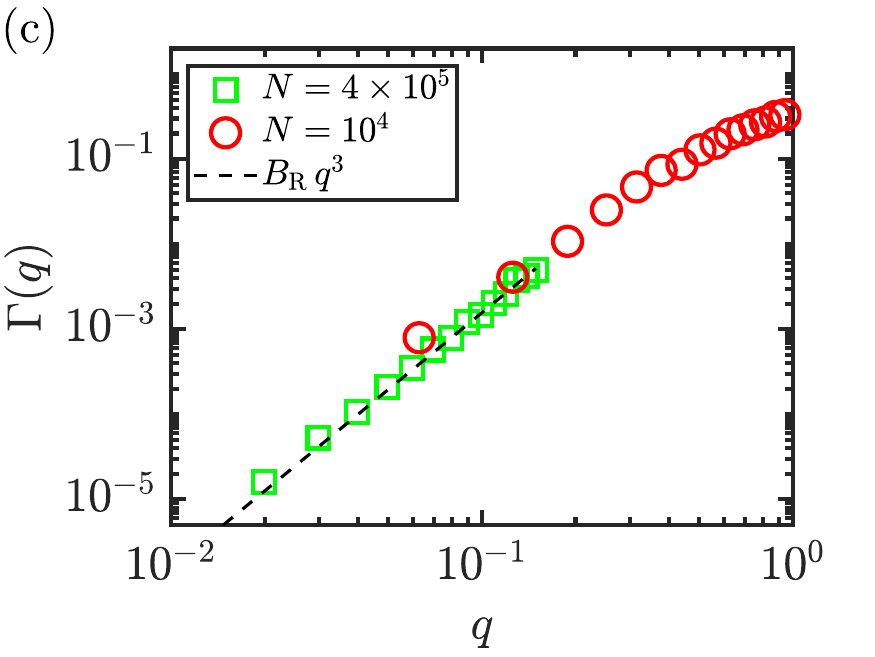}
    \end{subfigure}
    \begin{subfigure}[b]{0.4\textwidth}
     \includegraphics[width=1\textwidth]{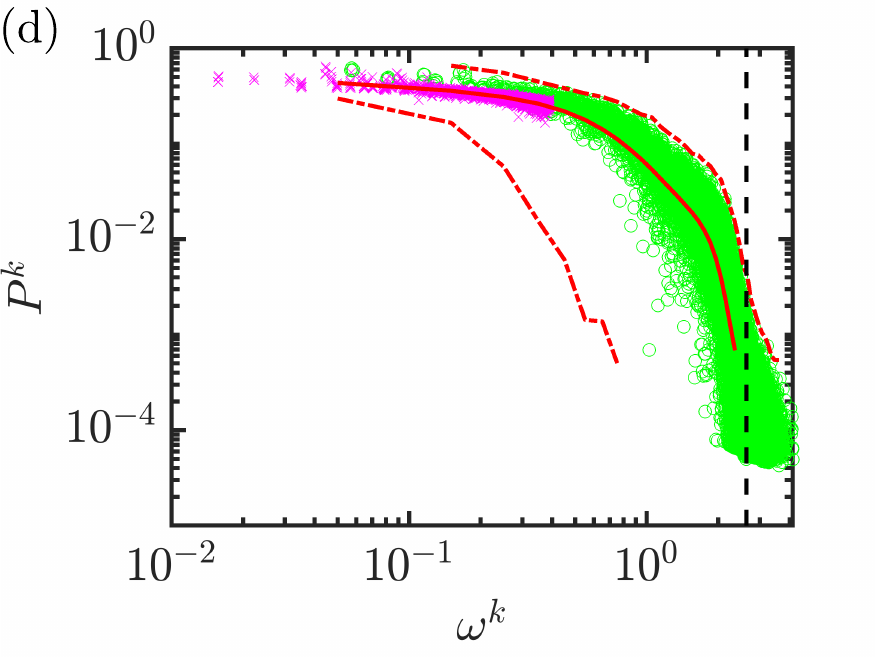}
    \end{subfigure}
    \caption{2D Systems. \textbf{(a)} Density of states $g(\omega)$ for  different densities $n$ divided by the Debye behaviour. \textbf{(b)} Obtained and rescaled Debye frequency $\hat{\omega}_\mathrm{D} =\omega_\mathrm{D}\, n^{-1}$ and boson peak frequency $\hat{\omega}_\mathrm{BP} =\omega_\mathrm{BP}\,n^{-1/2}$ in dependency of the density $n$. Additionally, the ratio $\hat{\omega}_\mathrm{BP}/\hat{\omega}_\mathrm{D}$. \textbf{(c)} Sound attenuation for the systems sizes $N = 10^4$ and $N = 4\times 10^5$ at $n=1$. A $q^3$ fit is shown for the small $q$ regime. \textbf{(d)} Participation ratio $P^k$ in dependency of the eigenfrequencies $\omega^k$ at $n=1$ for $N = 3\times 10^4$ (green \add{circles}) $N = 4\times 10^5$ (purple \add{crosses}). Only the 2000 lowest participation ratios can be calculated for $N =  4\times 10^5$. The black dashed line is located at $\omega_{BP}$. The red lines show the average, maximum and minimum participation ratio of an ensemble of systems with $N=3\times10^4$.}
    \label{fig:dos2D}
\end{figure*}
The contact number depends linear on the density $n$. This dependency becomes evident if one considers a $d$ dimensional sphere with the cut-off radius $\sigma$ and volume $V_d$ around a test particle. The contact number $z$ of the test particle  is given by the number of other particles in the sphere and thus simply reads $V_d\,n$. In Fig.~\ref{fig:contactNumber} the expected contact number  is shown by the dashed black lines.

At all densities $n$ the contact number is above Maxwell's isostatic stability criterion  $\overline{z}_c = 2\,d$ which is indicated by the dashed dotted lines \cite{Maxwell1864}. 
\add{We also checked the distribution of contact numbers, which also does not indicate a violation of Maxwell's criterion.}
Thus we conclude that the cut-off leads to negligible errors only. \add{We also confirmed this by calculations with cut-off $2\sigma$.}

\subsection{Two dimensional system}
We have also studied 2D systems. Compared to their $3D$ counterparts
the box size $L$ of the 2D systems at our maximum particle number of $N = 10^6$ is larger. Hence, smaller wavevectors $q$ are accessible in the 2D systems. In general, the 2D systems behave quite similar to their 3D counterparts. Note that the Debye behaviour becomes $g(\omega) = A_\mathrm{D}\,\omega$ in 2D with $A_D=1/\omega_D^2$.\\
In Fig.~\ref{fig:dos2D} the density of states $g(\omega)$, the dependences of the frequencies on density, $\hat{\omega}(n)$, the damping $\Gamma(q)$, and the participation ratios $P^k$ of the 2D systems are shown. We observe a Debye spectrum for $\omega \to 0$. Lowering the density $n$ again increases the Debye level i.e. decreases the Debye frequency. The Boson peak frequency $\omega_\mathrm{BP}$ again scales with $\sqrt{n}$, while the Debye frequency scales with $n^{1}$. We observe Rayleigh damping for small wavevectors $q$. Note that in 2D the Rayleigh damping becomes $\Gamma(q)\propto q^3$ \cite{Mizuno2018}\add{, as the fit in Fig.~\ref{fig:dos2D}(c) shows}. \\
In Fig.~\ref{fig:dos2D}(d) the corresponding participation ratios are shown. Again as in 3D, a crossover from extended to localized modes can be observed, which happens already at lower frequencies than $\omega_{BP}$ in 2D. An ensemble of $500$ systems allows to characterize the distribution of $P^k$, which lies lower at the boson peak frequency in 2D than in 3D; compare with Fig.~\ref{fig:dos2D}(d).

\section{Conclusions and outlook}\label{sec:conclusions}
Aim of the present contribution is to argue that the ERM system at large contact numbers captures pertinent phenomena of transverse vibrations in stable glass at vanishing temperature \cite{Wang2019, Wang2019-2, Mizuno2018,Monaco2009}. \add{It is thus a simple model which can be studied independent e.g.~on quenching protocols.} Two spatial dimensions, $d=3$ and $d=2$, were studied. We used dimensionless frequencies and wavevectors, which can be mapped onto experimental systems in the following way: The frequency or time scale of the ERM model is given by the boson peak frequency which was denoted $\omega_{BP}$. The corresponding length scale is obtained using the (transverse) sound velocity, $q_{BP}= \omega_{BP}/c_T$.  Both scales allow to map the vDOS and the dynamical structure factor onto measured data. The strength of the disorder then is the only free parameter. Within the ERM model with a Gaussian spring function it is quantified by the dimensionless density $n$. When comparing to real systems, the amplitude of the Debye-level at the position of the boson peak, viz.~$g_D(\omega_{BP})=A_D\, \omega_{BP}^2$ relative to the amplitude of the boson peak, viz.~$g(\omega_{BP})$,
 can be used to match  $n$. 
This gives the dominant variation with disorder in the ERM system, when rescaled variables
$\omega/\omega_{BP}$ and $q/q_{BP}$ are employed. Importantly, all parameters of the ERM model are thus accessible by experiment easily and by well defined procedures.
\add{As stated, the scalar ERM model should be applied only to transverse modes, and, considering the differences to particle simulations \cite{Mizuno2018,Wang2019, Wang2019}, only to frequencies below the boson peak. } 

Our numerical results should be compared to earlier numerical work on the same ERM system using diagonalizations and approximate calculations with the method of moments \cite{Martin-Mayor2001,Ciliberti2003}. We extend the range of accessible wavevectors so that a clear statement on the damping of sound modes becomes possible. Considering the slow crossover to $\Gamma(q)\propto q^{d+1}$ exhibited in Fig.~\ref{fig:damping} and panel (c) of Fig ~ \ref{fig:dos2D} for $q\to0$, we conclude that previously no statements on Rayleigh damping had been possible in the numerical ERM solutions.  \add{Our results prove Rayleigh damping over a large window in disorder. Additionally, we established the frequency windows where the eigenmodes are extended or localized.}
\add{We observed that the frequency window, where GOE random matrix modes dominate, moves strongly with $n$, viz.~with the strength of the disorder. For large density reaching down to $n\ge 1/4$, they form the boson peak of the vDOS which is given by Wigner's semicircle law.}

We also compared our numerical results to predictions obtained from two self-consistent theories, one
where all diagrams of first order in a diagrammatic perturbative expansion in $1/n$ were re-summed \cite{Grigera2001,Ciliberti2003}, and a more recent one, where all diagrams of second order in   $1/n$ were re-summed \cite{Vogel2023}. Both theories qualitatively agree in the predictions of a Wigner-semicircle law for the boson peak and of a Debye-law at low frequencies, with the crucial difference of the sound damping. Non-planar diagrams in the diagrammatic perturbation expansion are required to correctly predict Rayleigh damping of sound. The non-planar diagrams capture non-local correlations of elastic fluctuations and  arise first in second order in $1/n$. Thus, they were missed in the
older approach which consequently predicts hydrodynamic damping $\Gamma(q)\propto q^2$.  See Ref.~\cite{Vogel2023} for  more details on the comparison of both theories.  
 
 Finally, let us address the interpretation of the main peak in the vDOS of the ERM model. When discussing Fig.~\ref{fig:dos} and Fig.~\ref{fig:dos2}, we called this peak boson peak and suggested to consider it as simple model of the (transverse contribution to the) boson peak in the vDOS of real glasses of simple particle systems. This interpretation, which differs from the older literature on the ERM system  \cite{Grigera2001,Ciliberti2003,Schirmacher2019}, rests on the following arguments:\\
$(i)$ The boson peak in the vDOS of real glasses  survives in the zero temperature limit and thus arguably should be contained in an harmonic approach such as the ERM model.\\
$(ii)$ The  reported universality of the boson peak \cite{MarruzzoSchirmacher, tanguy2023vibrations} is  mirrored in its origin in the disorder, which is the single conceptual extension of the harmonic approach to disordered solids beyond the classical Born-Debye theory of crystals: The nature of the dominant normal modes changes in the frequency region of the boson peak. While the vibrational modes are mainly extended below $\omega_{BP}$, they are localised above of the peak \cite{Weak_Localization_1984, Universality_BP_PRL_2004, Schirmacher1998}, which can be clearly seen in Fig.~\ref{fig:participationRatio}. 
$(iii)$ It was argued in \cite{Universality_BP_PRL_2004, Schirmacher1998} that the normal modes in the region of the Boson peak follow the statistics of a Gaussian orthogonal random  matrix ensemble (GOE).  This is why the shape of the boson peak resembles   Wigner's semicircle law  which is  one of the hallmarks of a  GOE matrix \cite{anderson2010introduction}.  Clearly, our peak shown and analysed   in the figures \ref{fig:dos}, \ref{fig:dos2}, \ref{fig:participationRatio}, and \ref{fig:dos2D} is in accordance with this characterisation of the boson peak. 
 This also  hints at why the planar ERM model \cite{Ciliberti2003, Grigera2001} describes this part of the spectrum rather well, both quantitatively and qualitatively. It is known, that only the planar diagrams survive in the thermodynamic limit for GOE matrices \cite{anderson2010introduction}.  Again,  see \cite{Vogel2023} for a detailed comparison of the planar and non-planar self-consistent model.\\
$(iv)$ The density of states of a stable ERM system exhibits only a single peak as seen in simulations of stable glasses \cite{Wang2019-2,Mizuno2018}. \\
$(v)$ An alternative explanation for the origin of the prominent peak in the ERM model is, that it is a smeared out  van Hove singularity \cite{Origin_BP_Smeared_out_VH_2004_PRL_Elliott}. But this can not rationalise the observed transition from extended to localised  eigenmodes at $\omega_{BP}$. Almost all eigenmodes above  $\omega_{BP}$ are localized (for the considered simple ERM spring function \add{ and the discussed $n$-range}) and while the dispersion relation approaches $\omega_{BP}$ in the limit of $q\to\infty$, the number of extended modes in the boson peak region is far too small in order to justify its interpretation as smeared van Hove singularity.
This is further supported by the analytic derivation of Wigner's semicircle law. It rests on the solution of the self-consistency equation for $g(\omega)$ neglecting the coupling to other modes  \cite{Grigera2001,Vogel2023}. Again, this is in accordance with the observed  transition in the nature of the normal modes.  \\ 

The present study considered the simplest ERM system of a Gaussian spring function \add{at small disorder}. Work is under way to extend it to \add{stronger disorder and} richer spring functions. Additionally, recent works have shown that the vectorial character of the displacement field in glass is important as vortex like eigenmodes become possible  \cite{Mizuno2017,Lerner2021,Baggioli2021,Wu2023,tanguy2023vibrations}.  We expect that these structures can be studied in appropriately generalized  ERM systems. 
											
\section{Acknowledgments}		 We thank Giancarlo Ruocco, Walter Schirmacher, and Grzegorz Szamel  for fruitful discussions.   The work was supported  by the Deutsche Forschungsgemeinschaft (DFG) via SFB 1432 project CO7.

\bibliography{quellen.bib} 
\end{document}